\documentclass[pra,aps,twocolumn,showpacs]{revtex4}

\usepackage{graphicx}

\begin{document}

\title{Continuous-variable quantum process tomography with squeezed-state probes}

\author{Jarom\'{i}r Fiur\' a\v sek}
\affiliation{Department of Optics, Palack\'y University, 17. listopadu 1192/12,  771~46 Olomouc, Czech Republic}

\begin{abstract}
We propose a procedure for tomographic characterization of continuous variable quantum operations which employs homodyne detection and single-mode squeezed probe states 
with a fixed degree of squeezing and anti-squeezing and a variable displacement and orientation of squeezing ellipse. Density matrix elements of a quantum process 
 matrix in Fock basis can be estimated by averaging well behaved pattern functions over the homodyne data. We show that this approach can be 
 straightforwardly extended to characterization of quantum measurement devices. The probe states can be mixed, 
 which makes the proposed procedure feasible with current technology. 
\end{abstract}


\maketitle

\section{Introduction}

As the complexity of quantum information processing devices increases, there is a growing need for tools for their characterization and benchmarking.
Quantum operations and channels can be completely characterized by quantum process tomography \cite{Poyatos97,Chuang97,Luis1999,Luis00,Fiurasek01}, 
which represents an extension of quantum state tomography \cite{Leonhardt97,Paris04,Lvovsky09} to quantum operations. 
Typically, the quantum operation $\mathcal{E}$ is probed with a sufficient number of input states $\rho_j$, measurements in several different bases
are performed on the output states $\mathcal{E}(\rho_j)$, and the quantum operation is reconstructed from the experimental data. Alternatively, 
in the ancilla-assisted quantum process tomography \cite{DAriano01,Dur01,Altepeter03,DeMartini03,Mohseni08} the operation $\mathcal{E}$
is probed with one part of a single fixed entangled bipartite state $\rho_{AB}$ and $\mathcal{E}$  is determined from measurements on the output bipartite state.
This latter approach is based on the Choi-Jamiolkowski isomorphism \cite{Choi75,Jamiolkowski72}, which tells us that if the probe state is pure and maximally entangled, 
$|\Phi\rangle_{AB}=\frac{1}{\sqrt{d}}\sum_{j=0}^{d-1}|jj\rangle_{AB}$, then the output bipartite state  
$\chi_{AB}= \mathcal{I}_A\otimes \mathcal{E}_B(|\Phi\rangle\langle \Phi|)$ is directly isomorphic to the operation $\mathcal{E}$.  
Here $d$ denotes the dimension of input Hilbert space $\mathcal{H}_{\mathrm{in}}$ and the states $|j\rangle$ form an orthonormal basis in $\mathcal{H}_{\mathrm{in}}$.

Quantum process tomography  works particularly well 
for few-qubit systems, and it has been successfully applied in the past to characterization of various single-qubit 
and two-qubit operations \cite{Nielsen98,Childs01,Mitchell03,Altepeter03,DeMartini03,OBrien04,Riebe06,Cernoch08}. As the number of qubits $N$ increases,
the full tomography becomes challenging since the number of parameters that have to be estimated grows exponentially with $N$. 
In some cases, scalable quantum process reconstruction may be achieved 
e.g. by approximating the operator $\chi$ by a matrix product state \cite{Baumgratz13,Plenio14} or by using compressed sensing techniques \cite{Gross10,Shabani11,Rodionon14}. 

Besides the issue of Hilbert space dimension, 
the quantum process tomography is also affected by the range of practically accessible input probe states. 
This is particularly relevant for continuous variable quantum process tomography \cite{Luis1999,Luis00,Lobino08,Rahimi11,Anis12,Kumar13},
which aims at characterization of quantum operations on modes of quantized electromagnetic fields. Here, the most natural and readily available probe states are represented by 
coherent states $|\alpha\rangle$, and the output states can be conveniently measured with homodyne detectors \cite{Leonhardt97,Lvovsky09}. 
Recently, this approach has been successfully employed to characterize a single-mode lossy channel \cite{Lobino08}, 
and a conditional single-photon addition and subtraction \cite{Kumar13}. Moreover, the coherent states were also used 
as probes for complete tomographic characterization of single-photon detectors \cite{Lundeen09,Feito09,DAuria11,Brida12b,Humphreys15}.   
Probing quantum processes with coherent states essentially amounts to determining a Husimi $Q$-function of the operator $\chi$. 
More precisely, assuming that the measurements on output states are described by a POVM with elements $\Pi_j$, the probability of measurement 
outcome $\Pi_j$ for input probe coherent state $|\alpha\rangle$ reads  $p_j(\alpha)=\mathrm{Tr}[|\alpha^\ast\rangle\langle \alpha^\ast |\otimes \Pi_j \,\chi]$. 
Usually, one would like to reconstruct the matrix elements of $\chi$ in Fock basis. To see the connection between the Husimi $Q$-function and the matrix elements in Fock basis,
recall that the $Q$-function of an operator $A$ is defined as $Q(\alpha)=\langle \alpha |A| \alpha\rangle/\pi$. We have
\begin{equation}
Q(\alpha,\alpha^\ast)=\frac{e^{-|\alpha|^2}}{\pi} \sum_{m,n=0}^\infty \frac{\alpha^{\ast m} \alpha^{n}}{\sqrt{m!\,n!}} A_{m,n},
\label{Qfunction}
\end{equation} 
which shows that the $Q$-function is a generating function of matrix elements $A_{m,n}=\langle m|A|n\rangle$ in Fock basis \cite{Rahimi11},
\begin{equation}
A_{m,n}= \left.\frac{\pi}{\sqrt{m!\,n!}} \frac{\partial^{m+n}}{\partial\alpha^{\ast m} \partial \alpha^{n}} \left[Q(\alpha,\alpha^\ast)e^{|\alpha|^2}\right]\right|_{\alpha=\alpha^\ast=0}.
\label{Amn}
\end{equation}
Here $\alpha$ and $\alpha^\ast$ are formally treated as independent variables. Estimation of $A_{m,n}$ from experimental data requires inversion of Eq. (\ref{Qfunction}) 
when the $Q$-function is not known precisely. This can be a delicate procedure sensitive to statistical fluctuations of the data.

In Ref. \cite{Lobino08}, elements of quantum process matrix of a lossy channel were 
reconstructed from the experimental data with the help of regularized version of Glauber-Sudarshan $P$-functions \cite{Klauder66} of operators $|m\rangle\langle n|$. 
This approach approximates the calculation of derivatives in Eq. (\ref{Amn}) by evaluation of a suitable linear combination of the experimental data. 
In Ref. \cite{Lundeen09}, POVM elements of a single-photon detector were reconstructed from measurements on probe coherent states by solving a convex optimization 
problem that included an extra constraint which ensured a smooth structure of the reconstructed POVM elements. Later on, maximum-likelihood estimation was
employed for reconstruction of quantum operations and detectors probed with coherent states \cite{DAuria11,Kumar13}. This latter approach avoids the complications with direct 
linear  inversion (\ref{Amn}), but it requires some truncation of the infinite-dimensional operator $\chi$. 

In this paper, we investigate characterization of continuous variable quantum operations which is based on single-mode squeezed probe states and homodyne detection on output states. 
By using squeezed states instead of coherent states we avoid the problems with linear inversion of the data and we show that the matrix elements of quantum process $\chi$
in Fock basis can be determined by averaging suitable well behaved pattern functions \cite{DAriano95,Leonhardt95b,Richter00} over the homodyne data. 
Our procedure assumes that all probe states have the same variances of squeezed and anti-squeezed quadratures, 
and these variances need to be known and kept constant during the whole measurement. The probe states also need to be phase shifted and coherently
displaced in a controlled way, which is feasible with current technology. Importantly, our procedure works for realistic mixed squeezed states and the only requirement 
is that the variance of the squeezed quadrature is below the coherent state level. Although we focus on linear reconstruction based on the formalism of pattern functions,
the data could be processed by other means, such as the maximum likelihood estimation. Our work provides an important  insight into the utility 
of squeezed states for tomography of quantum processes.

\section{Quantum process tomography}

In what follows we shall consider chracterization of a single-mode quantum operation $\mathcal{E}$. According to the Choi-Jamiolkowski isomorphism \cite{Choi75,Jamiolkowski72},
such operation can be represented by a positive semidefinite operator $\chi$ on a Hilbert space of two modes,
\begin{equation}
\chi= \mathcal{I}\otimes \mathcal{E}(\Psi),
\label{chi}
\end{equation}
where $\mathcal{I}$ stands for the identity channel and  $\Psi=|\Psi\rangle \langle \Psi|$ denotes a density matrix of an infinitely squeezed EPR state,
\begin{equation}
|\Psi\rangle =\sum_{n=0}^\infty |nn\rangle.
\label{Psi}
\end{equation}
The input-output transformation $\rho_{\mathrm{out}}=\mathcal{E}(\rho_{\mathrm{in}})$ can be expressed as
\begin{equation}
\rho_{\mathrm{out}}=\mathrm{Tr}_{\mathrm{in}}\left[ \rho_{\mathrm{in}}^T \otimes I \,\chi\right],
\label{rhoout}
\end{equation}
where $T$ denotes transposition in Fock basis, $I$ stands for the identity operator, and $\mathrm{Tr}_{\mathrm{in}}$ denotes partial trace over the input mode.
In Fock basis, the formula (\ref{rhoout}) explicitly reads,
\begin{equation}
\rho_{\mathrm{out},m,n}=\sum_{k=0}^\infty \sum_{l=0}^\infty \chi_{km,ln}  \rho_{\mathrm{in},k,l}.
\label{rhooutmn}
\end{equation}
Here $\rho_{m,n}=\langle m|\rho|n\rangle$ and $\chi_{km,ln}=\langle km|\chi|ln\rangle$. Identity channel $\mathcal{I}$ is isomorphic
to the EPR state (\ref{Psi}), $\chi_\mathcal{I}=\Psi$, and $\chi_{\mathcal{I},km,ln} =\delta_{km}\delta_{ln}$.

The input-output transformation (\ref{rhoout}) can be also formulated for phase-space representations. Let $W_{\mathrm{in}}(x,p)$ and $W_{\mathrm{out}}(x,p)$ denote the Wigner
functions of input and output density operators $\rho_{\mathrm{in}}$ and $\rho_{\mathrm{out}}$, respectively, and let $W_{\chi}(x_{\mathrm{in}},p_{\mathrm{in}},x_{\mathrm{out}},p_{\mathrm{out}})$ 
denote the Wigner function of operator $\chi$. The partial trace (\ref{rhoout}) can be rewritten as an integral over the phase space of the input mode,
\begin{eqnarray*}
W_{\mathrm{out}}(x_{\mathrm{out}},p_{\mathrm{out}})&=&2\pi\int_{-\infty}^\infty \int_{-\infty}^\infty  W_{\mathrm{in}}(x_{\mathrm{in}},-p_{\mathrm{in}}) \\
& & \times W_{\chi}(x_{\mathrm{in}},p_{\mathrm{in}},x_{\mathrm{out}},p_{\mathrm{out}})d x_{\mathrm{in}} d p_{\mathrm{in}}.
\end{eqnarray*}
where $W_{\mathrm{in}}(x_{\mathrm{in}},-p_{\mathrm{in}})$ is a Wigner function of the transposed input state $\rho_{\mathrm{in}}^T$.

\begin{figure}[t]
\includegraphics[width=0.9\linewidth]{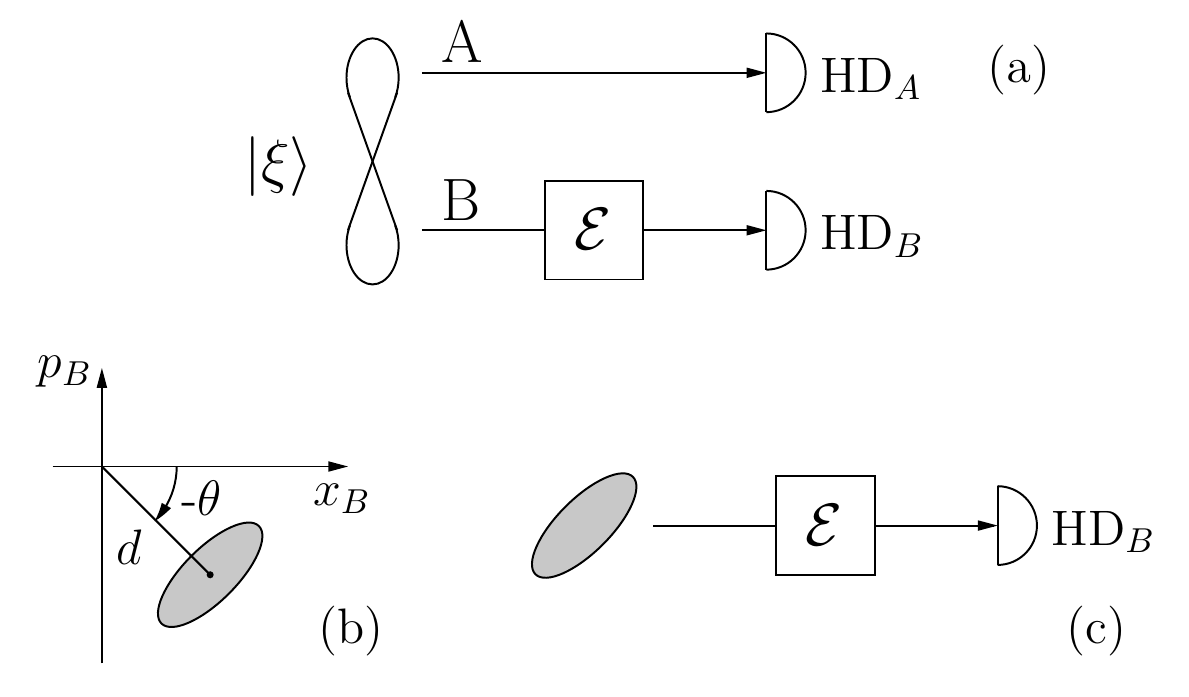}
\caption{(a) Ancilla-assisted quantum process tomography \cite{Luis1999,DAriano01,Dur01} of a single-mode operation $\mathcal{E}$. The operation is applied to one part of input two-mode squeezed vacuum state $|\xi\rangle$
and both output modes are measured with balanced homodyne detectors HD$_A$ and HD$_B$. (b) Homodyne measurement of quadrature $x_A^\theta$ of mode A of the two-mode squeezed 
vacuum state $|\xi\rangle$ prepares mode B in a coherently displaced and rotated
Gaussian squeezed state. (c) The ancilla assisted tomography is therefore equivalent to probing the operation $\mathcal{E}$ with a suitably chosen ensemble of single-mode squeezed states.}
\end{figure}

Our goal is to establish a procedure for determination of the matrix elements $\chi_{km,ln}$ from experimental data. 
Formula (\ref{chi}) suggests that this could be achieved by probing the quantum operation $\mathcal{E}$ with one part of the EPR state $|\Psi\rangle$.  
To make this continuous-variable ancilla-assisted quantum process tomography \cite{,DAriano01,Dur01}
experimentally feasible, the unphysical infinitely squeezed EPR state  may be replaced with a two-mode squeezed vacuum with finite squeezing \cite{Luis1999},
\begin{equation}
|\xi\rangle=\sqrt{1-\lambda^2}\sum_{n=0}^\infty \lambda^{n}|nn\rangle,
\label{xi}
\end{equation}
where $\lambda=\tanh r$, and $r$ denotes the squeezing constant.
After some algebra, we find that the elements of the quantum process matrix $\chi$ can be determined as properly rescaled elements of the
output two-mode state  $\sigma^\lambda=\mathcal{I}\otimes \mathcal{E}(\xi)$, where  $\xi=|\xi\rangle\langle \xi|$,
\begin{equation}
\chi_{km,ln}=(1-\lambda^2)^{-1}\lambda^{-(k+l)} \sigma^\lambda_{km,ln}.
\label{chielements}
\end{equation}

If both modes of the output state $\sigma^\lambda$ would be measured with homodyne detectors, see Fig.~1(a), then the matrix elements $\sigma^\lambda_{km,ln}$ 
could be reconstructed by quantum homodyne tomography \cite{Leonhardt97,Lvovsky09}. Let $\eta_A$ and $\eta_B$ denote the 
overall detection efficiency of balanced homodyne detectors HD$_A$ and HD$_B$, respectively. A detector with efficiency $\eta$ can be modeled 
as a lossy channel with transmittance $\eta$ followed by an ideal detector with unit efficiency. 
The detectors measure rotated quadratures of modes A and B, which are specified by angles $\theta$ and $\phi$, respectively,
\begin{eqnarray}
x_A^\theta&=&\sqrt{\eta_{A}}\left(x_A\cos\theta+p_{A}\sin\theta\right)+\sqrt{1-\eta_A}x_{A,\mathrm{vac}}^\theta, \nonumber \\
x_B^\phi&=&\sqrt{\eta_{B}}\left(x_B\cos\phi+p_{B}\sin\phi\right)+\sqrt{1-\eta_B}x_{B,\mathrm{vac}}^\phi. \nonumber \\
\end{eqnarray}
Here $x_J$ and $p_J$ denote the amplitude and phase quadratures of mode $J$, $[x_J,p_K]=i\delta_{JK}$, 
and $x_{A,\mathrm{vac}}^\theta$ and $x_{B,\mathrm{vac}}^\phi$ represent quadratures of auxiliary vacuum modes.
This measurement samples the joint
quadrature distribution $P(x_A^\theta,x_B^\phi;\theta,\eta_A,\phi,\eta_B)$ and the matrix elements $\chi_{km,ln}$ can be determined by averaging the so-called 
pattern functions over the quadrature statistics \cite{DAriano95,Leonhardt95b,Richter00},
\begin{widetext}
\begin{equation}
\chi_{km,ln}=\frac{(1-\lambda^2)^{-1}}{4\pi^2\lambda^{k+l}}\int\limits_{-\infty}^\infty \int\limits_{0}^{2\pi}\int\limits_{-\infty}^\infty\int\limits_{0}^{2\pi}
P\left(x_A^\theta,x_B^\phi;\theta,\eta_A,\phi,\eta_B\right) f_{k,l}(x_A^\theta,\eta_A)f_{m,n}(x_B^\phi,\eta_B) e^{i(k-l)\theta} e^{i(m-n)\phi} 
d x_A^\theta d\theta dx_B^\phi d\phi.
\end{equation}
\end{widetext}
Here $f_{m,n}(x,\eta)$ represent the loss-compensating single-mode pattern functions for density matrix elements in Fock basis. Explicit 
analytical expressions for $f_{m,n}(x,\eta)$ are provided in Ref. \cite{Richter00}. Since these expressions are rather cumbersome, we do not reproduce them here. 
We only note that the pattern functions $f_{m,n}(x,\eta)$ are well defined for $\eta>\frac{1}{2}$ and they diverge when $\eta \rightarrow \frac{1}{2}$.

\section{Single-mode probe states}

In this section, we will propose a procedure for quantum process tomography with single-mode squeezed probe states. In particular, we will exploit the fact that 
the ancilla assisted process tomography with a two-mode squeezed vacuum state $|\xi\rangle$ and individual single-mode homodyne measurements on the output modes is 
equivalent to preparation of a specific ensemble of displaced and rotated single-mode squeezed states of mode B, 
followed by probing the quantum operation $\mathcal{E}$ with these states \cite{Plenio14}. To see this equivalence, we rewrite the joint quadrature distribution as
\begin{eqnarray}
P(x_A^\theta,x_B^\phi;\theta,\eta_A,\phi,\eta_B)&= &P(x_A^\theta;\theta,\eta_A) \nonumber \\
& & \times P(x_B^\phi;\phi,\eta_B|x_A^\theta;\theta,\eta_A), \nonumber \\
\label{Pjoint}
\end{eqnarray}
where $P(x_A^\theta;\theta,\eta_A)$ is the probability density of measurement outcomes $x_A^\theta$ on mode A, and 
$P(x_B^\phi;\phi,\eta_B|x_A^\theta;\theta,\eta_A)$ is the conditional probability density of measurement outcomes of quadrature $x_B^\phi$ 
on mode B provided that a particular measurement outcome $x_A^\theta$ was obtained on mode A. 

Since mode A is in a thermal state, the probability $P(x_A^\theta;\theta,\eta_A)$ does not depend on $\theta$, and all quadratures $x_A^\theta$ exhibit Gaussian distribution with zero mean and 
\ variance 
\begin{equation}
V_A=\frac{1}{2}\left[\eta_A\cosh(2r)+1-\eta_A\right].
\end{equation}
This formula accounts for imperfect detection with efficiency $\eta_A$, and $\frac{1}{2}\cosh(2r)$ is the variance of quadratures of mode A of the
 pure two-mode squeezed vacuum state (\ref{xi}). Explicitly, the probability density reads
\begin{equation}
P\left(x_A^\theta;\theta,\eta_A\right)= \frac{1}{\sqrt{2\pi V_A}} \exp\left[ -\frac{\left(x_A^{\theta}\right)^2}{2V_A}\right].
\label{PxA}
\end{equation}

Homodyne detection of quadrature $x_A^\theta$ on mode A of the two-mode squeezed vacuum state (\ref{xi}) prepares the other mode B in a coherently displaced squeezed state
with squeezing ellipse rotated by angle $-\theta$, see Fig.~1(b). This rotation follows from the identity
\begin{equation}
U_A(\theta) U_B(-\theta) |\xi\rangle = |\xi\rangle,
\label{UAB}
\end{equation}
where $U(\theta)=e^{-in\theta}$ is a unitary phase shift operator. Measurement of a rotated quadrature $x_A^\theta$ on mode A is thus fully equivalent to measurement
of quadrature $x_A$, followed by rotation of mode B by $-\theta$. The covariance matrix of the conditionally prepared state 
does not depend on the measurement outcome $x_A^\theta$, and the coherent displacement $d$ is linearly proportional to the
measurement outcome. 

Let $V_{-}$ and $V_{+}$ denote the variances of squeezed and anti-squeezed quadratures of the conditionally prepared state, and let $d$ denote the coherent displacement 
od the squeezed quadrature of this state. It follows from the above discussion that, without loss of generality, we can assume $\theta=0$ in our derivation 
of  $V_{-}$, $V_{+}$, and $d$. It is convenient to collect the quadrature operators of modes A and B into a vector $z=(x_A,p_A,x_B,p_B)$  and define a two-mode covariance matrix 
$\gamma_{jk}=\langle\Delta z_j\Delta z_k+ \Delta z_k\Delta z_j\rangle$, where $\Delta z_j=z_j-\langle z_j\rangle$.
Covariance matrix of a two-mode squeezed vacuum state (\ref{xi}) whose mode A was transmitted through a lossy channel with transmittance $\eta_A$
reads,
\begin{equation}
\gamma_{AB}= \left(
\begin{array}{cccc}
2V_A & 0 & K & 0 \\
0 & 2V_A & 0 & -K \\
 K & 0 & 2V_B & 0 \\
 0 & -K & 0 & 2V_B
\end{array}
\right),
\end{equation}
where $V_B=\frac{1}{2}\cosh(2r)$ and $K=\sqrt{\eta_A}\sinh(2r)$.

Since there are no correlations between the $x_A$ and $p_B$ quadratures, measurement of $x_A$ does not influence $p_B$, whose variance remains equal 
to $V_B$ and $\langle p_B\rangle=0$,
\begin{equation}
V_{+}=\frac{1}{2}\cosh(2r).
\label{Vplus}
\end{equation}
In contrast, the measurement of $x_A$ will reduce fluctuations of $x_B$ due to the correlations between $x_A$ and $x_B$. The resulting (conditional) variance $V_{-}$ of $x_B$
can be calculated by minimizing the variance of $x_B-g x_A$ over a tunable gain $g$. The optimal gain reads $g_{\mathrm{opt}}=K/(2V_A)$,
which yields 
\begin{equation}
V_{-}=\frac{1}{2} \frac{\eta_A+(1-\eta_A)\cosh(2r)}{\eta_A\cosh(2r)+1-\eta_A}.
\label{Vminus}
\end{equation}
Moreover, the coherent displacement $d$ of the conditionally prepared state of mode B is given by $\langle x_B\rangle=g_{\mathrm{opt}}x_{A}$,
which explicitly reads
\begin{equation}
d=\frac{\sqrt{\eta_A}\sinh(2r)}{\eta_A\cosh(2r)+1-\eta_A} \,x_A.
\label{d}
\end{equation}

The variances $V_{-}$ and $V_{+}$ of squeezed and anti-squeezed quadratures of the probe single-mode state determine the effective detection efficiency $\eta_A$ of HD$_A$ 
and the parameter $\lambda$ of the (virtual) two-mode squeezed vacuum state (\ref{xi}). By inverting formulas (\ref{Vplus}) and (\ref{Vminus}), we get
\begin{equation}
\eta_A= \frac{2(V_{+}-V_{-})}{(2V_+-1)(2V_{-}+1)},
\label{etaA}
\end{equation}
and
\begin{equation}
\lambda=\sqrt{\frac{2V_{+}-1}{2V_{+}+1}}.
\end{equation}
The effective detection efficiency $\eta_{A}>\frac{1}{2}$ if and only if the probe state is squeezed and $V_{-}<\frac{1}{2}$.
This establishes single-mode squeezing as a valuable resource for continuous variable quantum process tomography.
If the probe state is pure, $V_{+}=1/(4V_{-})$, then $\eta_A=1$. If $V_{-}<\frac{1}{2}$ then the efficiency is a decreasing function of $V_{+}$ and in the limit 
$V_{+}\rightarrow \infty$ we get $\eta_A=1/(1+2V_{-})$.  The coherent displacement (\ref{d}) of mode B can be expressed in terms of the quadrature variances
as follows,
\begin{equation}
d= \sqrt{2(V_{+}-V_{-})}\sqrt{\frac{2V_{-}+1}{2V_{+}+1}} x_A.
\end{equation}

Formula (\ref{Pjoint}) together with the above results suggests that the joint quadrature distribution $P(x_A^\theta,x_B^\phi;\theta,\eta_A,\phi,\eta_B)$ 
can be sampled as follows. Generate random $x_A^\theta$ drawn from the Gaussian distribution (\ref{PxA}) and a random $\theta$ and $\phi$ drawn from a uniform distribution in the $[0,2\pi]$
interval.  Prepare a single-mode squeezed Gaussian state with variances $V_{-}$ and $V_{+}$ and displacement $d$, rotated in phase space by $-\theta$, as illustrated in Fig.~1(b). 
Send this probe state through the quantum channel $\mathcal{E}$ and measure a rotated quadrature $x_B^\phi$ of the output state with a homodyne detector. 
 Note that the squeezing properties of the input probe states do not depend on $x_{A}^\theta$ and $\theta$, 
 hence  a source producing squeezed states with a fixed amount of squeezing and anti-squeezing is sufficient. 
 
 The improvement achieved by squeezed probe states
 in comparison to coherent probe states comes at a cost of somewhat increased experimental difficulty. In particular, the variances $V_{+}$ and $V_{-}$
 of the probe squeezed states need to be precisely characterized, which can be achieved by routine homodyne detection, and these parameters
 have to be kept constant during the whole tomographic measurement. Moreover, the orientation of the squeezing ellipse should be fully under control and tunable to any required angle $\theta$.
 Finally, the ability to coherently displace the squeezed state is also required, which can be achieved e.g. by mixing it with an auxiliary coherent beam on a highly unbalanced beam splitter \cite{Furusawa98}.

\section{Tomography of quantum measurements}

The proposed method can be also adapted to tomographic characterization of quantum measurements \cite{Luis1999b,Fiurasek01b,Lundeen09,Feito09,DAuria11,Amri11,Brida12}.
 Consider a detector D which can respond with $K$ different outcomes.  Each outcome is associated with 
a POVM element $\Pi^{k}$ and the probability to observe an outcome $k$ for input state $\rho$ reads $p(k)=\mathrm{Tr}[\Pi^{k}\rho]$. 
Consider now an ancilla-assisted quantum detector tomography \cite{Amri11,Brida12}, where 
the detector is probed with one part of a two-mode squeezed vacuum state, see Fig.~2(a). 
The conditionally prepared state of mode A corresponding to measurement outcome $k$ on mode B can be expressed as
\begin{equation}
\rho^k=\left(1-\lambda^2\right) \sum_{m=0}^\infty \sum_{n=0}^\infty \lambda^{m+n} \Pi_{m,n}^k |n\rangle \langle m|.
\label{rhok}
\end{equation}
The state is not normalized and its trace is equal to probability of observing the outcome $k$, 
\begin{equation}
\mathrm{Tr}(\rho^k)= \langle \xi | I_A\otimes \Pi_B^k |\xi\rangle.
\end{equation}
Formula (\ref{rhok}) implies that the information about the POVM element $\Pi^k$  is imprinted into the conditional state $\rho^k$. In particular, we have
\begin{equation}
\Pi_{m,n}^k = \lambda^{-(m+n)}\left (1-\lambda^2\right)^{-1} \rho_{n,m}^k,
\label{Pimn}
\end{equation}
in analogy with Eq. (\ref{chielements}). The conditional states $\rho^k$ can be characterized by homodyne detection on mode A, 
which would provide sufficient data to reconstruct the density matrix elements $\rho_{m,n}^k$.

\begin{figure}[t]
\includegraphics[width=0.9\linewidth]{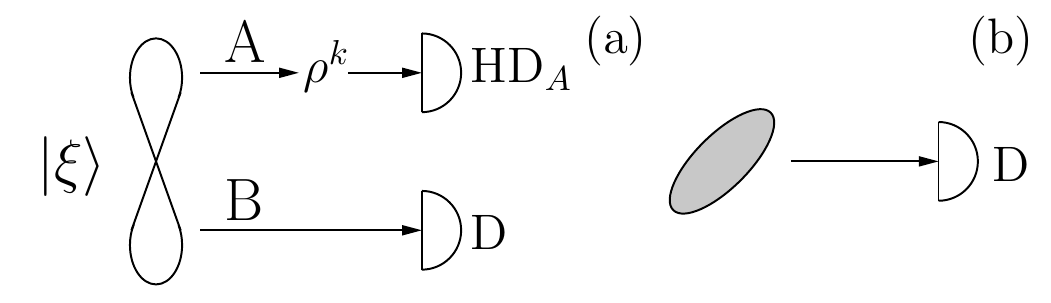}
\caption{(a) Ancilla-assisted quantum detector tomography \cite{Amri11,Brida12}. An unknown detector D is probed with one part of two-mode squeezed vacuum state $|\xi\rangle$. Information  about POVM element $\Pi^k$
associated with measurement outcome $k$ of detector $D$ is imprinted into the corresponding conditional state $\rho^k$ of mode A,
which can be characterized by homodyne tomography. (b) An equivalent scheme where the detector is probed with an ensemble of Gaussian squeezed states.}
\end{figure}

Similarly as for quantum operations, probing with one part of two-mode squeezed vacuum can be replaced by probing with single-mode squeezed states, see Fig.~2(b). 
Let $p(k|x_A^\theta,\theta)$ denote the probability of outcome $k$ for a probe state with displacement and rotation specified by parameters $x_A^\theta$ and $\theta$, 
respectively, c.f. Eqs. (\ref{Vplus}), (\ref{Vminus}), and (\ref{d}). The statistics of homodyne measurements on $\rho^k$ is governed by 
$P(x_A^\theta,\theta,\eta_A) p(k|x_A^\theta,\theta)$, where $\eta_A$ is a function of the variances of squeezed and anti-squeezed quadratures of the probe squeezed state, see Eq. (\ref{etaA}).
The density matrix elements of $\rho^k$ can be obtained by averaging appropriate pattern functions over the quadrature statistics,
\begin{eqnarray}
\rho_{m,n}^k &=&\frac{1}{2\pi}\int_{0}^{2\pi} \int_{-\infty}^\infty  P(x_A^\theta,\theta,\eta_A) p(k|x_A^\theta,\theta) \nonumber \\
& & \times f_{m,n}(x_A^\theta,\eta_A) e^{i(m-n)\theta} d x_A^\theta d\theta.
\end{eqnarray}
The matrix elements of $\Pi^k$ can then be immediately obtained from Eq. (\ref{Pimn}), 
where the  parameter $\lambda=\tanh r$ is determined by the variance of anti-squeezed quadrature $V_{+}$ of the probe state, see Eq. (\ref{Vplus}).

\section{Conclusions}

In summary, we have proposed a procedure for tomographic characterization of continuous variable quantum operations which employs homodyne detection and single-mode squeezed probe states 
with a fixed degree of squeezing and anti-squeezing and a variable displacement and orientation of squeezing ellipse. 
We have shown that the elements of quantum process matrix $\chi$ in Fock basis
can be estimated by averaging suitable pattern functions over the homodyne data. The pattern functions are well behaved provided that the probe state is squeezed and $V_-<\frac{1}{2}$. 
For the sake of simplicity, we have considered tomography of a single-mode operation $\mathcal{E}$. However, the method can be straightforwardly extended to multimode operations. 
For tomography of $N$-mode operation, one would have to use $N$ independent single-mode squeezed states and measure each output mode with an independent homodyne detector. 
While we have focused on 
linear reconstruction procedure based on pattern function formalism, other methods of data processing would be also possible. For instance, one may utilize 
the widely employed maximum-likelihood estimation,  or other approaches. 
Given its relative simplicity and practical feasibility, 
the present procedure is likely to find  applications in the characterization of continuous variable quantum operations and measurements. 

\acknowledgments

The research leading to these results has received funding from the EU FP7 under Grant Agreement No. 308803 (Project BRISQ2), co-financed by M\v{S}MT \v{C}R (7E13032).

\end{document}